\documentclass[final,letterpaper,5p]{elsarticle}
\usepackage{graphicx,color}
\usepackage{amssymb}
\usepackage{amsmath}
\usepackage{latexsym}
\usepackage{amsxtra}
\usepackage{amscd}
\usepackage{dcolumn}
\usepackage{bm}
\usepackage{hyperref}
\biboptions{numbers,sort&compress}
\graphicspath{}
\let\oldsqrt\sqrt 
\def\sqrt{\mathpalette\DHLhksqrt}
\def\DHLhksqrt#1#2{\setbox0=\hbox{$#1\oldsqrt{#2\,}$}\dimen0=\ht0
\advance\dimen0-0.2\ht0
\setbox2=\hbox{\vrule height\ht0 depth -\dimen0}%
{\box0\lower0.4pt\box2}}

\newcommand{\nuc}[2]{$^{#1}$#2}

\newcommand{\betv}{$B(E2)$ value}

\journal{Physics Letters B}

\begin{document}
\begin{frontmatter}

\title{First spectroscopy of \nuc{61}{Ti} and the transition to the Island of Inversion at $N=40$}

\author[ut,rnc]{K.~Wimmer}
\author[upad,infnpad]{F.~Recchia}
\author[upad,infnpad]{S.~M.~Lenzi}
\author[uper,infnper]{S.~Riccetto}
\author[ed]{T.~Davinson}
\author[cmu]{A.~Estrade}
\author[ed]{C.~J.~Griffin}
\author[rnc]{S.~Nishimura}
\author[iphc]{F.~Nowacki}
\author[rnc,vnu]{V.~Phong}
\author[uam]{A. Poves}
\author[rnc]{P.-A.~S\"oderstr\"om}
\author[kth]{O.~Aktas}
\author[liv,sa]{M.~Al-Aqeel}
\author[ut]{T.~Ando}
\author[rnc]{H.~Baba}
\author[snu,inpa]{S.~Bae}
\author[snu,inpa]{S.~Choi}
\author[rnc]{P.~Doornenbal}
\author[snu,rnc]{J.~Ha}
\author[liv]{L.~Harkness-Brennan}
\author[rnc]{T.~Isobe}
\author[upad,infnpad]{P.~R.~John\corref{pre1}}
\author[ed]{D.~Kahl}
\author[rnc,ato]{G.~Kiss}
\author[gsi]{I.~Kojouharov}
\author[gsi]{N.~Kurz}
\author[dar]{M.~Labiche}
\author[ut]{K.~Matsui}
\author[ut]{S.~Momiyama}
\author[leg]{D.R.~Napoli}
\author[ut]{M.~Niikura}
\author[buc]{C.~Nita}
\author[rnc]{Y.~Saito}
\author[ut,rnc]{H.~Sakurai}
\author[gsi]{H.~Schaffner}
\author[cns]{P.~Schrock}
\author[dar]{C.~Stahl}
\author[rnc]{T.~Sumikama}
\author[dar]{V.~Werner}
\author[dar,gsi]{W.~Witt}
\author[ed]{P.~J.~Woods}

\address[ut]{Department of Physics, The University of Tokyo, 7-3-1 Hongo, Bunkyo-ku, Tokyo 113-0033, Japan}
\address[rnc]{RIKEN Nishina Center, 2-1 Hirosawa, Wako, Saitama 351-0198, Japan}
\address[upad]{Dipartimento di Fisica e Astronomia dell' Universit\`a di Padova, Padova I-35131, Italy}
\address[infnpad]{INFN, Sezione di Padova, Padova I-35131, Italy}
\address[uper]{Dipartimento di Fisica e Geologia dell' Universit\`a di Perugia, Perugia, Italy}
\address[infnper]{INFN, Sezione di Perugia, Perugia, Italy}
\address[ed]{School of Physics and Astronomy, The University of Edinburgh, James Clerk Maxwell Building,  Edinburgh EH9 3FD, United Kingdom}
\address[cmu]{Central Michigan University, USA}
\address[iphc]{IPHC, IN2P3-CNRS et Universit\'e de Strasbourg, F-67037 Strasbourg, France}
\address[vnu]{Faculty of Physics, VNU University of Science, 334 Nguyen Trai, Thanh Xuan, Hanoi, Vietnam}
\address[uam]{Department of Theoretical Physics and IFT UAM-CSIC Universidad Aut\'onoma de Madrid 28049, Madrid, Spain}
\address[kth]{KTH Royal Institute of Technology, Stockholm, Sweden}
\address[snu]{Department of Physics and Astronomy, Seoul National University, Seoul, 08826, Republic of Korea}
\address[inpa]{Institute for Nuclear and Particle Astrophysics, Seoul National University, Seoul, 08826, Republic of Korea}
\address[liv]{Department of Physics, University of Liverpool, Oliver Lodge Building, Oxford Street, L697ZE, Liverpool, UK}
\address[sa]{Department of Physics, College of Science, Al-Imam Mohammad Ibn Saud Islamic University (IMISU), Riyadh, 11623, Saudi Arabia}
\address[ato]{Institute for Nuclear Research (MTA ATOMKI), H-4001 Debrecen, POB.51., Hungary}
\address[gsi]{GSI Helmholtzzentum f\"ur Schwerionenforschung GmbH, Darmstadt, Germany}
\address[dar]{STFC Daresbury Laboratory, UK}
\address[leg]{INFN, Laboratori Nazionali di Legnaro, Legnaro (Padova), Italy}
\address[buc]{Horia Hulubei National Institute for Physics and Nuclear Engineering (IFIN-HH), Bucharest, Romania}
\address[cns]{Center for Nuclear Study, University of Tokyo, Hongo, Bunkyo-ku, Tokyo 113-0033, Japan}
\address[dar]{Institut f\"ur Kernphysik, Technische Universit\"at Darmstadt, Germany}

\cortext[pre1]{present address: Institut f\"ur Kernphysik, Technische Universit\"at Darmstadt, Germany}

\begin{abstract}
  Isomeric states in \nuc{59,61}{Ti} have been populated in the projectile fragmentation of a 345~$A$MeV \nuc{238}{U} beam at the Radioactive Isotope Beam Factory. The decay lifetimes and delayed $\gamma$-ray transitions were measured with the EURICA array. Besides the known isomeric state in \nuc{59}{Ti}, two isomeric states in \nuc{61}{Ti} are observed for the first time. Based on the measured lifetimes, transition multipolarities as well as tentative spins and parities are assigned. Large-scale shell model calculations based on the modified LNPS interaction show that both \nuc{59}{Ti} and \nuc{61}{Ti} belong to the Island of Inversion at $N=40$ with ground state configurations dominated by particle-hole excitations to the $g_{9/2}$ and $d_{5/2}$ orbits.
\end{abstract}

\date{\today}
\end{frontmatter}

\section{Introduction}
The study of nuclei far off stability in regions that have not yet been explored -- and in particular its description in terms of nuclear shell structure -- is fundamental in order to obtain predictive capabilities and to be able to understand the relevant characteristics of such nuclei. Due to enormous computational advancements in the recent years and the development of new interactions, the shell model is now able to predict the structure of nuclei far from closed shells. One example of this are the neutron-rich $N=40$ nuclei below \nuc{68}{Ni} ($Z=28$). 
Although the energy and the \betv\ of the first $2^+_1$ state of \nuc{68}{Ni} suggest a shell closure, the neutron separation energies show a smooth trend across $N=40$~\cite{sorlin02}. By removing a few protons, a rapid increase of collectivity has been observed in the Fe and Cr isotopes, suggesting a weakening of the $N=40$ sub-shell closure~\cite{rother11,crawford13}.

This collective behavior is caused by quadrupole correlations which energetically favor the deformed intruder states involving the neutron $1g_{9/2}$ and $2d_{5/2}$ orbitals and proton excitations across the $Z=28$ sub-shell gap. Moreover, the imbalance between protons and neutrons modifies the shell structure in this region. These changes in the intrinsic shell structure are of fundamental interest for testing the validity of modern interactions and their predictive power farther from stability.
The subtle interplay between such shell-evolution mechanisms provokes the modification of the magic numbers and gives rise to new regions of deformation and shape coexistence phenomena. Previous studies indicate \nuc{64}{Cr} is the center of a new region of prolate deformation. As in the case of \nuc{32}{Mg} at $N = 20$, shape coexistence should be expected in this region~\cite{lenzi10}. 

The region of neutron-rich nuclei around $N=40$ is also characterized by the occurrence of long-lived excited states in these nuclei. The large difference in angular momentum between the $\nu 2p_{1/2}$, $1f_{5/2}$ and $2d_{5/2}$, $1g_{9/2}$ orbitals around the Fermi surface in $N\sim 40$ nuclei leads to the occurrence of isomeric states. These isomeric states are well established along the Ni and Fe isotopic chains~\cite{grzywacz98,block08,olaizola13}. Farther below in the Cr and Ti isotopes, such isomers should also exist. In \nuc{59}{Ti} an isomeric state with a half-life of 590~ns was discovered~\cite{kameda12}. Based on Weisskopf estimates, an $E2$ transition connecting the $5/2^-$ and $1/2^-$ states was proposed. A positive parity isomeric state based on the $\nu g_{9/2}$ intruder configuration is also expected. The location of such pure configurations in the most neutron-rich nuclei is paramount to test the predictions of shell model calculations for the most exotic $N=40$ nuclei \nuc{60}{Ca} and \nuc{59}{K} which are still beyond the reach of presently available radioactive beam facilities for spectroscopic studies. First experimental hints have been obtained from the first spectroscopy of \nuc{60}{Ti}~\cite{gade14} and the discovery of particle-bound \nuc{60}{Ca}~\cite{tarasov18}. 

In this letter, we present a search for new isomers in \nuc{59}{Ti} and report on the first spectroscopic study of \nuc{61}{Ti}. The experimental results are interpreted in terms of neutron holes for the $N=39$ nuclei and as excitations across the weak $N=40$ sub-shell closure. The excitation energies of these simple configurations provide direct insight in the structure, in an even more interesting and direct way than from the spectroscopy of the neighboring even-even nuclei.

\section{Experimental setup}
\begin{figure}[h]
\centering
\includegraphics[width=\columnwidth]{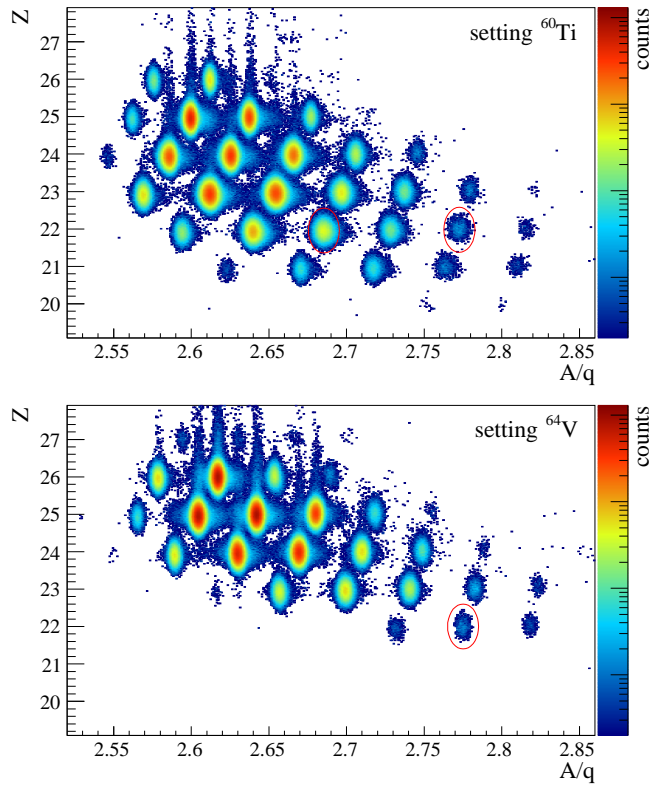}
\caption{Particle identification plot for (top) the \nuc{60}{Ti} setting, (bottom) the \nuc{64}{V} setting. The mass to charge ratio $A/q$ and the nuclear charge $Z$ are determined from the magnetic rigidity, time-of-flight, and energy loss of particles in BigRIPS.}
\label{fig:pid}
\end{figure}

The experiment was performed at Radioactive Isotope Beam Factory (RIBF), operated by RIKEN Nishina Center and CNS, University of Tokyo. A wide range of exotic neutron-rich nuclei were produced by fragmentation of a \nuc{238}{U} at 345~$A$MeV on a Be primary target (thickness 4~mm) at the entrance of the BigRIPS fragment separator~\cite{kubo12}. An average primary beam intensity of 23~pnA was achieved. In order to purify the beam, two wedge-shaped degraders with central thickness of 6 and 2.5~mm were located at the dispersive F1 and F5 focal planes of BigRIPS. Fragments were identified in $Z$ and $A/q$ by measuring the energy loss ($\Delta E$), time-of-flight ($TOF$), and the magnetic rigidity ($B\rho$) in BigRIPS ($\Delta E-TOF-B\rho$ method) and then transported to the experimental setup at the final F11 focal point of the ZeroDegree spectrometer. Data were taken in two settings, centered on \nuc{64}{V} and \nuc{60}{Ti}, respectively. 
The particle identification plots are shown in Fig.~\ref{fig:pid}.
The nuclei were then further degraded in energy and implanted in the Advanced Implantation Detector Array (AIDA)~\cite{griffin15}. The present experimental results are part of a $\beta$-decay experiment whose results will be presented in a forthcoming publication~\cite{riccetto18}. Here, we only concentrate on the decay of isomeric states, and thus AIDA served as a passive stopper. The array was surrounded by the high-purity Ge EUroball RIKEN Cluster Array (EURICA)~\cite{soederstroem13}, consisting of 84 Ge crystals with an efficiency of about 10\% at 1332~keV. Data were recorded independently by the BigRIPS, AIDA, and EURICA data acquisition systems, and correlated off-line by means of their synchronized time-stamps. This allowed for the analysis of short ($\sim \mu$s lifetime) and long lived isomeric decays, as well as $\beta$ decays.

\section{Experimental results}
\subsection{Isomeric decay of \nuc{59}{Ti}}
\begin{figure}[h]
\centering
\includegraphics[width=\columnwidth]{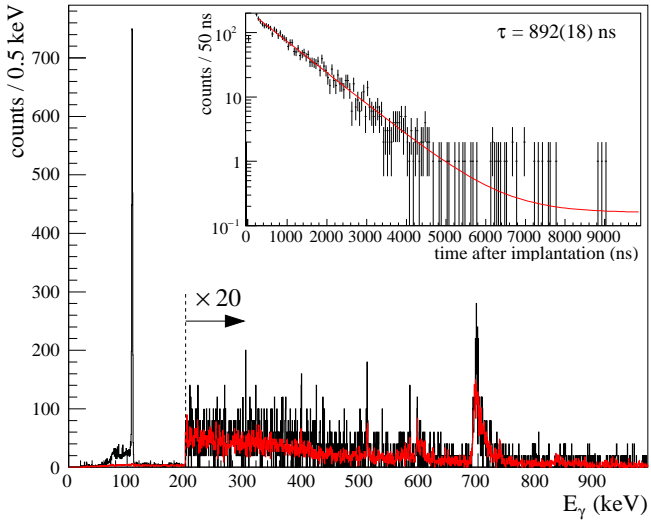}
\caption{Isomeric $\gamma$-rays detected within 10~$\mu s$ after the implantation of \nuc{59}{Ti}. The inset shows the decay curve for the 108.5~keV transition. The spectrum is also compared with \nuc{61}{V} (red, scaled for the number of implanted ions), which does not have an isomeric state, to show the neutron induced background around 700~keV. No other isomeric transition besides the 108.5~keV line is observed within this time window.}
\label{fig:ti59}
\end{figure}

The isomeric decay of \nuc{59}{Ti} was studied previously at RIBF~\cite{kameda12}. In the present study, a factor of $\sim\!20$ times more statistics was obtained. The delayed coincidence $\gamma$-ray energy spectrum is shown in Fig.~\ref{fig:ti59}. One transition at 108.5(5)~keV was observed.
The lifetime of the isomer was obtained from a decay curve fit to the $\gamma$-ray detection time after the implantation. The resulting lifetime is $\tau=892(18)$~ns. This value is in agreement with the previous measurement~\cite{kameda12}, but more precise due to the improved statistics.
No other delayed transition in \nuc{59}{Ti} has been observed based on the comparison with the neighboring isotopes. The background is well reproduced by scaling the corresponding spectrum for \nuc{61}{V} with no known isomeric state for the number of implanted ions.

Previous $\beta$-decay experiments~\cite{liddick05,crawford10} suggested a spin and parity of $J^\pi = 5/2^-$ for the ground state of \nuc{57}{Ti} based on systematics and comparison with shell model calculations using the GXPF1 interaction~\cite{honma02}. These calculations predict the first excited state with $J^\pi =1/2^-$ at 422~keV. Experimentally the first excited state is located at 364~keV~\cite{crawford10} and $J^\pi =1/2^-$ was assigned. However, the strong feeding of the excited state in the $\beta$ decay of the $J^\pi=7/2^-$ ground state of \nuc{57}{Sc} would suggest a reversed ordering. For \nuc{59}{Ti}, $J^\pi = (5/2^-)$ was assigned to the ground state as well~\cite{kameda12} based on systematics, and $(1/2^-)$ was proposed for the isomer. In the present work, using a total conversion coefficient $\alpha = 0.249$~\cite{kibedi08}, the transition probability for an $E2$ transition with $\tau=892(18)$~ns amounts to $B(E2) = 48.8(10)$~e$^2$fm$^4$ or $3.58(7)$~W.u.. Other assumptions for the multipolarity result in unreasonable transition probabilities. The excited state is assigned as $J^\pi = (1/2^-)$, if the ground state is $5/2^-$. From the present experimental data, it can not be excluded that the 108.5~keV $E2$ transition may originate from the decay of positive parity $9/2^+$ or $5/2^+$ states. In the \nuc{63,65}{Fe} nuclei the $9/2^+$ states are $\beta$-decaying isomers, thus longer correlation times (up to 10~ms) between the implantation of \nuc{59}{Ti} and subsequent detection of $\gamma$ rays were investigated. No other isomeric transition was found, in particular none in coincidence with the 108~keV transition. This indicates that the 108~keV transition is originating from the only $\gamma$-decaying isomer in \nuc{59}{Ti} and that it is a ground-state transition. If isomeric, other states like the positive parity $9/2^+$ or $5/2^+$ states could decay by $\beta$ emission. The $\beta$ decay of \nuc{59}{Ti} was investigated and the resulting lifetime is in agreement with previous work~\cite{daugas11} with only one $\beta$-decaying state.

\subsection{First spectroscopy of \nuc{61}{Ti}}
\begin{figure}[h]
\centering
\includegraphics[width=\columnwidth]{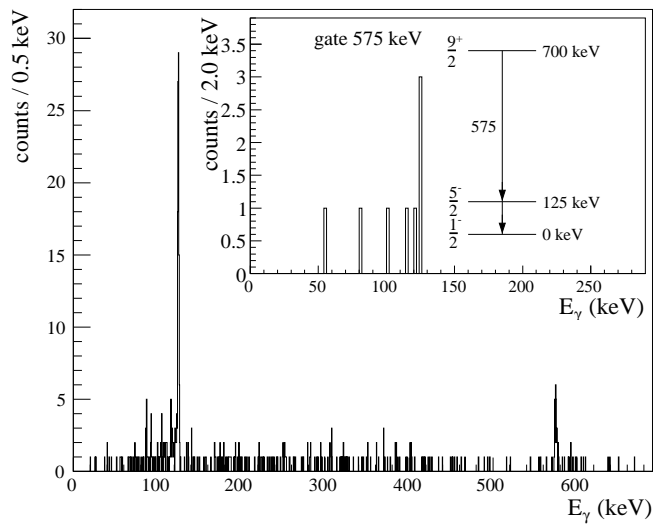}
\caption{Isomeric $\gamma$-rays for \nuc{61}{Ti}. The inset shows coincidences with the 125~keV transition and the proposed level scheme.}
\label{fig:ti61}
\end{figure}

\begin{figure}[h]
\centering
\includegraphics[width=\columnwidth]{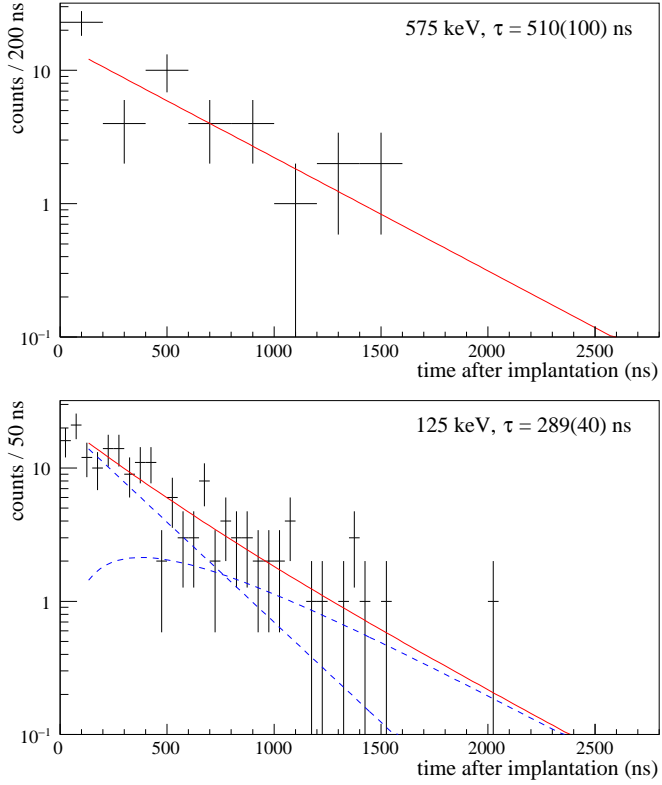}
\caption{Isomeric decay curves for \nuc{61}{Ti}. Time of the $\gamma$-ray detection after the implantation for the 575~keV (top) transition fitted with an exponential decay curve. The 125~keV transition is fed by this transition, and thus its decay curve (bottom) has been fitted taking into account the lifetime of the feeding transition. The red line shows the total decay curve, while dashed blue lines show the two components of the Bateman equations.}
\label{fig:ti61_tau}
\end{figure}

In \nuc{61}{Ti} two delayed transitions at 125.0(5) and 575.1(5)~keV were observed. The $\gamma$-ray energy spectrum is shown in Fig.~\ref{fig:ti61}.
$\gamma-\gamma$ coincidences show that the two transitions originate from a cascade decay of a level at 700~keV. Based on the intensity and the lifetimes (see below) the existence of two isomeric states in \nuc{61}{Ti} are suggested, both with approximately equal ratios in the beam at the implantation position. 
The level at 700~keV decays to the 125~keV state by the 575~keV transition; however, indirect feeding of the 125~keV state alone does not explain the number of counts observed in the spectrum in Fig.~\ref{fig:ti61}. The full-energy peak detection efficiency of EURICA in the present configuration amounts to $~\approx 12$\% at 575~keV and $\approx 25$\% at 125~keV, while about four times as many counts are observed in the 125~keV transition in Fig.~\ref{fig:ti61}. To corroborate this assumption, the lifetimes of the two states were determined in Fig.~\ref{fig:ti61_tau}.
Firstly, the lifetime of the 700~keV state was determined by gating on the 575~keV transition. The resulting value of $510(100)$~ns was subsequently used in the Bateman equation for the decay curve of the 125~keV state. The lifetime of $289(40)$~ns of the 125~keV state was found to be rather insensitive to the value of the feeding lifetime. As in the case of \nuc{59}{Ti}, the lifetime suggests an $E2$ transition with $B(E2) = 81(10)$~e$^2$fm$^4$ or $5.7(8)$~W.u.. For the 575~keV transition, the only reasonable choice of multipolarity is $M2$ with $B(M2) = 2.3(5)$~$\mu^2$fm$^2$ or $0.09(2)$~W.u.. The parity of the 700~keV state is thus positive. This fixes the spin and parity of the 700~keV state to be $J^\pi = 9/2^+$ and the one of the lower isomer at 125~keV to be $J^\pi = 5/2^-$ (otherwise decay to the ground state would be more probable). The ground state is then assigned $J^\pi = (1/2^-)$ based on the $E2$ transition. The suggested level scheme is shown in the inset of Fig.~\ref{fig:ti61}. 

This result also puts a constraint on the location of the $5/2^+$ state in \nuc{61}{Ti}. Assuming 1~W.u. for the transition of the $9/2^+$ state to the $5/2^+$ state and a partial lifetime that is 10 times longer than the $\tau(M2;\;9/2^+ \rightarrow 5/2^-)$ measured here, such that the decay is not observable, the $5/2^+$ state can only be located 96~keV below the $9/2^+$ state or at any energy higher. For less conservative assumptions, the limit is more stringent. 

\section{Discussion}
\begin{figure}[htb]
\centering
\includegraphics[width=\columnwidth]{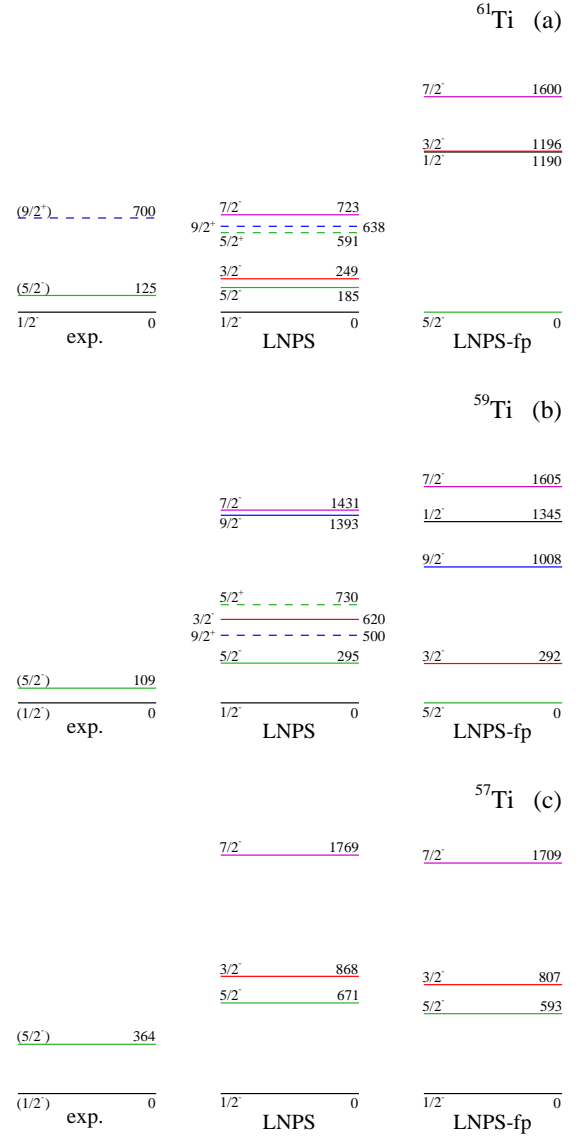}
\caption{Level schemes of \nuc{61-57}{Ti} in comparison with large scale shell model calculations using the LNPS effective interaction. LNPS-fp denotes the LNPS interaction with neutrons constrained to the full $pf$ model space. See text for details.}
\label{fig:levelsti}
\end{figure}

The present data have been interpreted in the framework of the large-scale shell model calculations using the LNPS interaction~\cite{lenzi10}. The model space consists of the full $pf$ shell for protons and the $1f_{5/2}$, $2p_{3/2}$, $2p_{1/2}$, $1g_{9/2}$, and $2d_{5/2}$ neutron orbitals outside a \nuc{48}{Ca} core.
The results are shown in Fig.~\ref{fig:levelsti} in comparison with the present data. Calculations with the same interaction but limited to the full $pf$ shell for both protons and neutrons, i.e. using a \nuc{40}{Ca} core and not allowing excitations to the $gd$ orbits, are denoted as LNPS-fp.  
The calculations for \nuc{61}{Ti} in the large model space accurately reproduce the data. The ground state is predicted to be $J^\pi = 1/2^-$, with an excited first $5/2^-$ state at 185~keV. Using the effective charges $e_{\pi}=1.31$ and $e_{\nu}=0.46$~\cite{dufour96}, the calculated transition probability for the decay to the ground state is $B(E2;\;5/2^-\rightarrow 1/2^-)=60$~e$^2$fm$^4$, compatible with $81(10)$~e$^2$fm$^4$ determined from the measured lifetime. The wave functions are dominated by $\nu (fp)^{-5}(gd)^4$ configurations. The clear intruder dominance in the ground state shows that \nuc{61}{Ti} belongs to the Island of Inversion. Regarding the positive parity states, the lowest state is a $5/2^+$ state at 591~keV, while the $9/2^+$ state, lying just 47~keV above it, becomes isomeric. The calculations are compatible with the data and therefore $J^\pi = (1/2^-)$ is assigned to the ground state, $J^\pi = (5/2^-)$ to the first state at 125~keV, and $J^\pi = (9/2^+)$ to the 700-keV isomeric state in \nuc{61}{Ti}. A very different level scheme is predicted if the model space is limited to the $pf$ shell as can be seen in right side of Fig.~\ref{fig:levelsti} (a). No isomers are predicted in this case as the $J^\pi=5/2^-$ ground state is connected to any of the excited states by fast $E2$ or $M1$ transitions.

For $^{59}$Ti, in the large model space the spin of the ground state is predicted to be $J^\pi = 1/2^-$ (as for \nuc{61}{Ti}), with a first $J^\pi = 5/2^-$ excited state at 295~keV. The calculated $B(E2;\;5/2^-\rightarrow 1/2^-)$ value is 14~e$^2$fm$^4$.
The configuration of the wave function of the ground state is dominated by two neutrons in the $gd$ orbitals, which indicates that \nuc{59}{Ti} also belongs to the Island of Inversion around $N=40$. This is confirmed by the fact that in the $pf$ model space, the level scheme shown in Fig.~\ref{fig:levelsti} (b) is incompatible with the experimental findings.
Based on the good agreement of the calculations with the LNPS interaction with the data we propose to assign $J^\pi = (1/2^-)$ to the ground state and $J^\pi = (5/2^-)$ to the isomer in \nuc{59}{Ti}. The first $9/2^+$ state is predicted at 500~keV and the $5/2^+$ state is located 230~keV above it at 730~keV. This suggests a $9/2^+$ isomeric state. Since it is located below the $5/2^+$ state, it can only decay via a $M2$ transition to the first excited state. Experimentally only one isomer has been observed in the present work. However, as stated above, the possibility of another long-lived state can not be ruled out. If the lifetime of such a state is significantly shorter than the flight time through the separator, or it is a $\gamma$-decaying isomer with a lifetime longer than $\sim10$~ms, or if such an isomer was not populated in the fragmentation reaction, it would not have been observed in the present experiment. 

It is interesting to explore the evolution of the structure in the lighter Ti isotopes to investigate the boundaries of the Island of Inversion along the isotopic chain.
In \nuc{57}{Ti} at $N=35$, only two states with suggested negative parity are known experimentally~\cite{crawford10}. Originally the ground state had been assigned $J^\pi =5/2^-$ with an excited $1/2^-$ state based on comparison with the GXPF1 interaction~\cite{honma02}. On the contrary, as shown in Fig.~\ref{fig:levelsti} (c), the calculations with the LNPS interaction in both model spaces predict a ground state with $J^\pi = 1/2^-$ and a first excited $5/2^-$ state. This spin sequence also agrees with the observed large feeding in the $\beta$ decay of the $J^\pi = 7/2^-$ ground state of \nuc{57}{Sc} to the excited state at 364~keV in \nuc{57}{Ti}. This suggests a change of the spin assignments proposed in \ref{fig:levelsti} (c). The ground state of \nuc{57}{Ti} is dominated in the LNPS calculations by $fp$ configurations, with marginal excitations to the $gd$ orbitals, which locates therefore $N=35$ \nuc{57}{Ti} outside the $N=40$ Island of Inversion. Similar conclusions can be drown for \nuc{55}{Ti}. 

The success in describing the data for these heavy Ti isotopes is encouraging for predictions above the $N=40$ sub-shell closure. For \nuc{63}{Ti} a similar spectrum as in \nuc{61}{Ti} is expected with the $5/2^-$ state at 76~keV excitation energy and the $3/2^-$ state at 140~keV, while the $7/2^-$ remains higher in energy at 620 keV. The positive-parity states suggest the occurence of only one isomeric state in this nucleus. At $N=40$ and removing two protons, \nuc{60}{Ca} is predicted to have ground, $2^+$ and $4^+$ yrast states with dominant $4p-4h$ neutron excitations into the $1g_{9/2}$ and $2d_{5/2}$ orbits~\cite{lenzi10}.
Looking at the $N=39$ isotones from Ni to Ca, the calculations predict $0p-0h$ excitations into the $gd$ orbitals for the ground state of \nuc{67}{Ni}, $2p-2h$ for \nuc{65}{Fe}, and $4p-4h$ for \nuc{63}{Cr}, \nuc{61}{Ti} and \nuc{59}{Ca}. In determining the collectivity and $B(E2)$ values, the proton configurations play a major role. As for the $N=40$ isotones~\cite{lenzi10}, maximum collectivity is obtained mid-shell for \nuc{63}{Cr}, with a decreasing trend towards $Z=20$. 

In summary, first spectroscopic information of \nuc{61}{Ti} has been obtained through isomeric spectroscopy. Two states were observed, suggesting two isomeric states, a $9/2^+$ state at 700~keV and a $5/2^-$ state at 125~keV. The lifetimes were measured and from comparison to Weisskopf estimates the transition multipolarities and thus spin and parity assignments were made. Candidates for positive parity states in \nuc{59}{Ti} were not observed, but they might have been out of the reach of the present experiment. The results suggest that particle-hole excitations across the $N=40$ sub-shell closure play a major role in the wave function of the low-lying states of neutron-rich Ti isotopes. Both \nuc{59}{Ti} and \nuc{61}{Ti} belong to the Island of Inversion around $N=40$.

\section{Acknowledgments}
This experiment was carried out at the RIBF operated by RIKEN Nishina Center, RIKEN and CNS, University of Tokyo. We would like to thank the RIKEN accelerator and BigRIPS teams for providing the high intensity beams. 
This work has been supported by JSPS KAKENHI (Grant No. 25247045), by the German BMBF (Grant No. 05P15RDFN1), by the STFC (UK), by the Korean National Research Foundation (grants NRF-21A20131111123, NRF-2015H1A2A1030275), by NKFIH (NN128072), and by the UNKP-18-4-DE-449 New National Excellence Program of the Ministry of Human Capacities of Hungary. G.~Kiss acknowledges support from the Janos Bolyai research fellowship of the Hungarian Academy of Sciences. A. Poves acknowledges support by Mineco (Spain) grants FPA2014-57916 and Severo Ochoa Program SEV-2016-0597.
\section{References}
\bibliographystyle{elsarticle-num-names}
\bibliography{draft}

\end{document}